\documentclass[proof]{pasj00}

\begin{document}
\SetRunningHead{Author(s) in page-head}{Mayama et al.2005}

\title{SUBARU Near-Infrared Multi-Color Images of Class~II Young Stellar Object, RNO91}


\author{Satoshi \textsc{Mayama,}\altaffilmark{1,2}
   Motohide \textsc{Tamura,}\altaffilmark{1,3}
   Masahiko \textsc{Hayashi,}\altaffilmark{1,2}
   \\
   Yoichi \textsc{Itoh,}\altaffilmark{4}
   Miki \textsc{Ishii,}\altaffilmark{2}
   Misato \textsc{Fukagawa,}\altaffilmark{5}
   \\
   Saeko S. \textsc{Hayashi,}\altaffilmark{1,2}
   Yumiko \textsc{Oasa,}\altaffilmark{4}
   and
   Tomoyuki \textsc{Kudo,}\altaffilmark{1,3}

   }

\altaffiltext{1}{School of Mathematical and Physical Science, Graduate University for Advanced Studies,\\ 650 North A'ohoku Place, Hilo, HI96720, U.S.A.}
\email{mayamast@subaru.naoj.org}

\altaffiltext{2}{SUBARU Telescope, National Astronomical Observatory of Japan,\\650 North A'ohoku Place, Hilo, HI96720, U.S.A.}

\altaffiltext{3}{Optical and Infrared Astronomy Division, National Astronomical Observatory of Japan,\\2-21-1 Osawa, Mitaka, Tokyo 181-8588}

\altaffiltext{4}{Graduate School of Science and Technology, Kobe University, \\1-1 Rokkodai, Nada, Kobe 657-8501}

\altaffiltext{5}{Division of Particle and Astrophysical Sciences, Nagoya University, \\Furo-cho, Chikusa-ku, Nagoya 464-8602}


%

\KeyWords{stars:individual (RNO91)---stars:pre-main sequence---ISM:reflection nebula} 

\maketitle

\begin{abstract}

We conducted sub-arcsecond near-infrared imaging observations of RNO91 with CIAO mounted on the SUBARU 8.2 m telescope.  We present our $JHK$ band data along with optical images, which when considered together reveal a complex circumstellar structure.  We examined the colors of associated nebula and compared the geometry of the outflow/disk system suggested by our data with that already proposed on the basis of previous studies.  Our $K$-band image shows bright circumstellar nebulosity detected within $\sim$2$''$~(300AU) around the central source while it is less conspicuous at shorter wavelengths such as $J$ and optical.  P.A. and size of this red color nebulosity in our $H-K$ color image agree with those of the previously detected polarization disk.  These data agreement indicate that this bright circumstellar nebulosity region which follows the reddening law might be attributed to a disk-like structure.  At $J$ and optical wavelengths, several blue knot-like structures are detected around and beyond the bright circumstellar nebulosity.  We suggest that these knotty reflection nebulae may represent disintegrating fragments of an infalling envelope.  The three-color composite image has an appearance of arc-shaped nebulosity extending to the north and to the east through the central source.  On the other end of this arc-shaped structure, the nebula appears to become more extended (2\rlap.{$''$}3 long) to the southwest.  We interpret these structures as roots of bipolar cavities opening to the northeast and southwest.  The complex distribution of reflection nebulosity seen around RNO91 appears to confirm the interpretation of this source as an object dispersing its molecular envelope while transitioning from protostar to T Tauri star.

\end{abstract}

\section{Introduction}

RNO91 is located in the L43 dark cloud in Ophiuchus at a distance of 160 pc (Herbst \& Warner 1981).@L43 is a typical site of low-mass star formation.  The cloud houses at least two young nebulous stars (RNO91 and RNO90, Cohen 1980).  RNO91 is an embedded source identified as a T Tauri star (Andr\'e \& Montmerle 1994, Levreault 1988).  It is classified as a M0.5 type T Tauri star and has a mass of 0.5~M$_\odot$ (Leverault 1988).  RNO91 lies within a parabolic-shaped region of lower extinction with a higher surface brightness rim than elsewhere in the L43 cloud, which Mathieu et al. (1988) referred to as the ``bay''.  The bay opens to the south while RNO91 lies near the closed end to the north.  

RNO91 is known as a source of complex molecular outflows (Myers et al. 1988; Mathieu et al. 1988; Parker et al. 1988).  The molecular outflow shows spatially separated redshifted and blueshifted lobes.  The redshifted outflow lobe extends to the north, while the blueshifted outflow lobe extends to the south (Levreault 1988; Myers et al. 1988; Mathieu et al. 1988).  Low velocities of the outflow, only a few km/s from systemic velocity assuming an outflow inclination angle of 20-30$^{\circ}$ to the plane of the sky, are difficult to reconcile with either jet or wind model (Lee et al. 2002).  CO J~=~2-1 observations toward the outflow show a weak low-velocity emission along the outflow axis at about 500$''$ away from the source (Bence et al. 1998).  Lee \& Ho (2005) presented maps of the integrated CO J~=~1-0 emission which also showed that southeast lobe appeared as a U-shaped shell extending $\sim$500$''$ to the southeast.  Inside the southeast lobe, CO structures consisting of a chain of knotty structures are seen along the axis (Lee \& Ho 2005).  Recent high resolution radio observations revealed the inner region within 30$''$ of the source and showed the presence of compact outflows (Arce \& Sargent 2006).  

Previous near-infrared and optical observations suggested that RNO91 is associated with a reflection nebula (Schild et al. 1989; Hodapp 1994; Weintraub et al. 1994) and shock-excited H$_2$ emission (Kumar et al. 1999).  Previous radio observations revealed that RNO91 is associated with an ammonia envelope (Mathieu et al. 1988), HCO$^{+}$ and N$_2$H$^{+}$ envelopes (Lee \& Ho 2005).

Optical and infrared photometry (from $U$ band to $L$ band) was obtained by Myers et al. (1987).  Schild, Weir, \& Mathier (1989) presented optical images of this object and noted that the extended structure in their images is aligned perpendicular to the outflow and might be attributed to material associated with a disk-like structure.  This disk-like structure seems to be extended about 1000 - 2000AU.  Scarrott, Draper, \& Tadhunter (1993) presented optical polarization map of the small-scale nebula surrounding RNO91 and also suggested the presence of a small-scale a few arcsec circumstellar disk, which has collimated the outflow leading to the visible nebula.  Weintraub et al. (1994) demonstrated with their $K$-band polarimetric image that RNO91 is surrounded by an edge-on circumstellar disk-like structure of radius 1700AU.  However, the geometry of RNO91, especially its inner region ($\sim$20$''$) is not well confirmed yet. 
RNO91, even though in the late stages of star formation, is still active in producing an outflow.  An outflow plays an important role in the star formation process through their strong physical impact on the environments of the young stellar objects.  Thus, the L43/RNO91 system is one of the best targets for studying these interactions such that how an outflow may reduce and disperse an infalling envelope.  High resolution imaging is needed to better understand the geometry of RNO91 in the most inner region and the interaction between outflow and disk.  Furthermore, RNO91 is an important candidate because it is considered to be a YSO in a transition phase between protostar and visible pre-main-sequence star (Parker 1991).

In this paper, we present the sub-arcsecond near-infrared and optical images of RNO91 obtained with the SUBARU telescope, and discuss the morphology of circumstellar structures associated with RNO91.
Observations and data reduction procedures are described in sections 2 and 3, respectively. 
Results and discussion are presented in sections 4 and 5, respectively.  
Section 6 summarizes the conclusions.

\section{Observations}

The infrared observations were carried out on 2001 May 15 with the Coronagraphic Imager with Adaptive Optics, CIAO (Tamura et al. 2000, Murakawa et al. 2004) mounted at the f/12 Cassegrain focus of the SUBARU 8.2~m telescope.  The pixel scale and the orientation of the 1024$\times$1024 InSb array (ALADDIN~II) were measured using the Trapezium (Simon et al. 1999) with the same optical configurations for the RNO91 observations and were 0\rlap.{$''$}0213$\pm$0\rlap.{$''$}0003 pixel$^{-1}$ (Itoh et al. 2005) and 5.7$^{\circ}\pm$0.8$^\circ$, respectively.
The field of view was 22$''\times$22$''$.  
The imaging observations were carried out in the $J$, $H$ and $K$ bands centered at 1.25, 1.65, and 2.2~$\mu$m, respectively.  The sky was clear and photometric and seeing was stable.  Adaptive optics was not used due to an unavailability of appropriate natural guide stars.  An occulting mask was not used.  The PSFs of the point source observed in the same night have sizes of 0\rlap.{$''$}4$\pm$0\rlap.{$''$}1 (FWHM).

We obtained 6 frames with 10~s $\times$ 3~co-adds, 10~s $\times$ 3~co-adds and 5~s $\times$ 6~co-adds for the $J$, $H$ and $K$ bands, respectively.  The photometric standard star FS141 (Hawarden et al. 2001) data and sky frames were obtained after the observations of RNO91.  Dome flats and dark frames were taken at the end of the night.

  Additional infrared observations were carried out on 2005 June 27 with CIAO.  
The average natural seeing was 0\rlap.{$''$}5.  The pixel scale and position angle of the detector were 0\rlap.{$''$}0213$\pm$0\rlap.{$''$}0001 pixel$^{-1}$ and 1.68$^\circ$$\pm$0.27$^\circ$ (Mayama et al. 2006).
We obtained 40, 80 and 3 frames for RNO91 with 10~s $\times$ 1~co-add, 1~s $\times$ 1~co-adds and 0.33~s $\times$ 40~co-adds for the $J$, $H$ and $K$ bands, respectively.  No photometric standard star was observed.
Dome flats and dark frames were taken at the end of the night.

  Hubble Space Telescope, $HST$ archive data which was already calibrated was also used.  The observations were carried out on 1999 January 21 with the WFPC2 (Wide Field Planetary Camera 2).  WFPC2 consists of Wide Field camera (WF3) and Planetary camera (PC1).  RNO91 was imaged with F606W filter.  The pixel scales of WF3 and PC1 are 0\rlap.{$''$}1 pixel$^{-1}$ and 0\rlap.{$''$}046 pixel$^{-1}$, respectively.  The field of view of WF3 and PC1 are 2.5$'\times$2.5$'$ (L-shaped) and 35$''\times$35$''$, respectively.  The total integration times of WF3 and PC1 are 300~s and 100~s, respectively.

\section{Data Reduction}

The Image Reduction and Analysis Facility (IRAF) software was used for all the data reduction.  
A dark frame was subtracted from each object frame, which was then divided by the dome-flat.  
Hot and bad pixels were removed and sky frame was subtracted.
We combined the frames of each band to produce the final images.

The $J$, $H$, and $K$ band images of RNO91 are presented in Figure~1 together with an $HST$-WFPC2 (WF3) optical image.  The $J$ band image is from our second observations (in 2005), while the $H$ and $K$ band images are from our first observations (in 2001).  Instrumental magnitude was used to calculate the value [mJy/arcsec$^{2}$] for $J$ band image in Figure~1 as no photometric standard star was observed during our observations in 2005.
Identifiers of each structural feature are noted in Figure 4.

Changing band filters of CIAO during an observation introduces a slight displacement of object's position on the detector.  Measuring precise positions of the central source is essential in comparing the structures amongst different wavelength images such as making color composite images or identifying structures.  However, it is hard to precisely measure the positions of the central source which is deeply embedded in circumstellar material or spatially elongated, such as RNO91.  Furthermore, in the RNO91 images, there are no background point sources to which we can refer.  Although there is a peak intensity position that we are able to measure, their positions could differ in each different band.

In order to overcome this problem, we adopted the following method to correct the displacement in position on the detector at different wavelength images.  First, the peak position of the point source that was observed at $J$, $H$, and $K$ in the same night with RNO91 were measured.  Second, we examined the displacement in positions of the peaks at different wavelengths.  Between the $K$ band and the $H$ band, derived displacements in position on a detector were 0.45$\pm$0.76 [pixels] at E-W direction and 0.48$\pm$0.14 [pixels] at N-S direction.  Between the $H$ band and the $J$ band, derived displacements in position on a detector were 0.11$\pm$0.78 [pixels] at E-W direction and 2.4$\pm$0.76 [pixels] at N-S direction.  Third, we used these values to shift the images of RNO91 in order to correct the position displacement on a detector.  Finally, the shifted images are used to make color composite images (Figure~2).

We lastly address the method of position registration between SUBARU and HST images.  In order to make three-color composite image, the structural feature identified as V in Figure 4 is used for position registration due to the following two reasons.  (i)~There is no back ground star which can be used for position registration between SUBARU and HST images.  (ii)~The peak positions of V are located at same position on $J$, $H$, and $K$ band images whose position displacements are already corrected by the method explained in the above paragraph.  (iii)~The SUBARU $J$ band image is very similar to the HST optical image.  These facts enable us to assume that the position of V does not differ among wavelengths.  A position uncertainty between the bands is stated in section 4.1.\\

\section{Results}

\subsection{Colors of RNO91}

Figure~2 shows $H-K$ color image of RNO91.  In this figure, strong emission indicates red region, while weak emission indicates blue region.  Notable color features in this figure are as follows.  The color within a distance of $\sim$2$''$ from the photo center of the nebula appears to be red ($H-K$ \textgreater 1.1), while the color beyond a distance of $\sim$2$''$ from the photo center of the nebula appears to be blue ($H-K$ \textless 0.7).  Interpretation of these colors is described in section 5.1.  Figure~3 shows three-color composite image of SUBARU-CIAO(K), SUBARU-CIAO(J), and HST-WFPC2(F606W) broadband observations of RNO91.  Note that Figure~3 is composed of images from multiple epochs separated by years.  The CIAO $K$-band is registered as red, CIAO $J$-band as green, and WFPC2 F606W as blue.  Note also that our $H-K$ color image (Figure~2) and three-color composite image (Figure~3) have a position uncertainty of up to 0\rlap.{$''$}02 and 0\rlap.{$''$}1 due to the issue as discussed in the previous section, respectively.  However, these uncertainties are small enough to discuss morphology of circumstellar structure surrounding RNO91.  In Section 5, we explain the details and interpret the features of Figures 2 and 3.

\subsection{Radial Surface Brightness Profile}

Figure~5 shows radial profiles of the surface brightness of RNO91 associated nebula with respect to the $K$-band peak position.  Data obtained in 2001 were used for all $J$, $H$, and $K$ band profiles in Figure~5 since no photometric standard star was observed during our 2005 observations.  RNO91 is comprised of several clumps or knot-like structures.  In order to eliminate the effect of such local clump structures, we examined radial surface brightness only along lines of P.A.=0, 30, 300, 330$^{\circ}$.  The surface brightness decreases as r$^{-2.1\pm0.2}$, r$^{-2.3\pm0.2}$, and r$^{-2.5\pm0.2}$ with the radius r from 48 to 300 AU, for $J$, $H$, and $K$ bands, respectively.  The surface brightness decreases as r$^{-0.5\pm0.2}$, r$^{-0.6\pm0.2}$, and r$^{-0.8\pm0.2}$ with the radius r from 300 to 688 AU, for $J$, $H$, and $K$ band, respectively.  The power-law dependence revealed the following two points.
(1) At all wavelengths, the profile showed a change of slope at $\sim$2$''$~(300AU) from the central peak, suggesting the morphological difference beyond this point.  
(2) The power-law dependence at $J$, $H$, and $K$ did not show a significant difference among wavelengths, indicating that the same scatter and absorption are occurring at all wavelengths regardless of dust sizes.

\section{Discussion}

The complex circumstellar structure was detected in the most inner region ($\sim$20$''$) as shown in Figures 1 and 3.  Figure 6 is a schematic description of the morphology of the RNO91.

\subsection{Circumstellar Disk-like Structure and Envelope}

The $K$-band image shows bright circumstellar nebulosity detected within $\sim$2$''$ around the central source while less nebulosity is seen at shorter wavelengths such as $J$ and optical (see Figure~1).  The nebula appears to become more isotropic with increasing wavelength.  Figure~2 shows much red color particularly in northern part of this bright circumstellar nebulosity detected within $\sim$2$''$ around the central source.  Examined $H-K$ and $J-K$ colors around this northern part of bright circumstellar nebulosity are about 1.1$\sim$1.2 and 2.9$\sim$3.2, respectively.  Our derived $H-K$ and $J-K$ colors for this part can well be explained by reddening law with assuming an intrinsic color of an M0.5 star.

  Here, we would like to interpret that this bright circumstellar nebulosity region might be attributed to a disk-like structure based on following two reasons.
(1) Scarrott, Draper, \& Tadhunter (1993) presented polarization maps of the RNO91 nebulosity and showed evidence for a small-scale (a few arcsec) circumstellar disk running south-east to north-west.  Our $H-K$ color map shows a reddish region ($H-K$ \textgreater 1.1, Figure~2) which extends northwest along the P.A. of the putative polarization disk.  Therefore, the direction of this red color region in our $H-K$ image is consistent with P.A. of their polarization disk.
(2) The radial surface brightness profile (Figure~5) shows a change of slope at $\sim$2$''$~ from the photo center of the nebula as mentioned in the previous chapter, quantitatively demonstrating a morphological difference beyond this point.  In terms of color, an inner region (\textless 2$''$) has much redder color than an outer region (\textgreater 2$''$).  The change in slope of the radial surface profile at a distance of $\sim$2$''$ from the photo center of the nebula appears to agree with the size scale (a few arcsec) of the previously detected polarization disk (Scarrott, Draper, \& Tadhunter 1993).  These data agreements indicate that this bright circumstellar nebulosity region which follows the reddening law might be attributed to a disk-like structure.  In order to confirm this matter, however, further observations in radio wavelength are needed to obtain velocity distribution data of this most inner circumstellar structure.

At $J$ and optical wavelengths, several newly found blue knot-like structures (M1, M2, N1, N2, and N3 in Figure~4) are detected around and beyond the bright circumstellar nebulosity.  Schild, Weir, \& Mathier (1989) conducted spectroscopic observations toward RNO91 and did not detect any emission lines in their spectra of the nebular features except for the bright H$\alpha$ also seen in RNO91.  They concluded that the nebular is mostly, and likely entirely, reflection light from RNO91, and not from other emission sources.  Positions of these blue knots are measured in our images and compared with those in images obtained by Schild, Weir, \& Mathier (1989).  These positions are almost consistent, and our images showed spatially well resolved knots.  Thus as suggested by Schild, Weir, \& Mathier (1989), these blue knots seen in our shorter wavelength images, are not other emission sources, but part of reflection nebula.  

Tendency of scattered light to be blue (Sellgren et al. 1992) can well explain our derived colors ($H-K$ \textless 0.7, Figure~2) for the knots and reflection nebula.  Furthermore, these blue knots seen in our images are spread over a region comparable in size to the envelope detected in HCO$^{+}$ J~=~1-0 emission ($\sim$10$''$) surrounding RNO91 (Lee \& Ho 2005).  It is thus natural to interpret that these knotty reflection nebulae located beyond 2$''$ from central source position may represent disintegrating fragments of an infalling envelope, as postulated by Mathieu et al. (1988) and Lee \& Ho (2005).  We are looking at a site how outflow disrupts an envelope in a transition phase YSOs.  Mathieu et al. (1988) suggested that the molecular outflow is disrupting the dense core in L43, and it may well be that the clumps are part of this erosion.  Lee \& Ho (2005) also detected clumps or knots-like structures around the central peak and interpreted them as fragments or part of erosion occurred in envelope structures which are disrupted by molecular outflow.  These previous studies support our interpretation.\\

\subsection{Outflow and Cavity}

Figure 3 has an appearance of arc-shaped nebulosity extending to the north and to the east through the central source (Q1, Q2, and T in Figure~4).  The northern ridge is 11$''$ long and eastern ridge is 7$''$ long.  On the other end of this arc-shaped structure, the nebula appears to become more extended (2\rlap.{$''$}3 long) to the southwest (P.A.$\sim$225$^{\circ}$) from the central source position in $J$ band and $HST$ optical images (see Figure~1 and L in Figure~4) then branching to two lobes (P and G in Figure~4) with apparent cavity.  Regarding the inner part within 2 to 3$''$ from central source position, our $H-K$ color (Figure~2) and three-color composite image (Figure~3) shows that only this south-west direction has the blue ($H-K$ \textless 0.6) extended feature while the other directions has much red ($H-K$ \textgreater 1.1) circular structure which we interpreted as a disk-like structure in previous section.  This color distribution indicates that the extinction at south-west direction from the central source position is less than the counterpart.  This extinction structure requires the geometry that south-west direction is the near side.  This also requires that south-west outflow needs to be the near side.  Therefore, we interpret these whole structures as roots of bipolar cavity opening to the southwest (near side) and northeast (far side) directions.

  There are four arguments in previous papers which support our suggested geometry.  First, Scarrott, Draper, \& Tadhunter (1993) detected a polarization disk which has collimated the outflow leading to the visible nebula.  Their derived orientation of polarization disk was 145$^{\circ}$, and an outflow axis which is collimated by this disk needs to be perpendicular to this orientation.  Hence, an orientation of extended nebula in our $J$ band and $HST$ optical images (marked as L \& G in Figure~4) is almost consistent with that of blueshifted outflow axis derived from Scarrott, Draper, \& Tadhunter (1993).  
  
  Second, Lee \& Ho (2005) mapped N$_2$H$^{+}$ J~=~1-0 emission and showed there is more N$_2$H$^{+}$ emission extending to northeast, while less N$_2$H$^{+}$ emission extending to southwest (Figure~1(C) in Lee \& Ho 2005).  They suggested that the scarcity of the N$_2$H$^{+}$ emission to the southwest was likely due to less material there.  As N$_2$H$^{+}$ mainly traces the dense and cool material in the envelope, this N$_2$H$^{+}$ distribution indirectly suggests a tendency of blueshifted molecular outflow sweeping away material located around southwest direction of the central source.  
  
  Third, Kumar, Anandarao, \& Davis (1999) reported the detection of the H$_2$ v~=~1-0 S(1) line at 2.122~$\mu$m and found the shock-excited H$_2$ emission originating in an outflow.  In their image, H$_2$ emission is clearly extended more in the southern direction than in the northern direction (Kumar et al. 1999).  Therefore, the direction of outflow seen in H$_2$ emission is consistent with the nebulosity seen at our short wavelengths (labeled L in Figure~4), indicating that blueshifted outflow has components extending to south or southwest.
  
  Fourth, several previous studies presented CO emission maps which probes the low-density, high-velocity outflow gas originating from RNO91.   Here, we focus on several recent data which show relatively \textit{small scale} outflow structures comparable to the size of our NIR images.  Arce \& Sargent (2006) presented $^{12}$CO (1-0) mapping in the inner most region with high spatial resolutions (4 to 6$''$), which is sensitive to the bright, compact scale outflow structures within 30$''$ from the source.  In their $^{12}$CO (J=1-0) contour maps, blueshifted emission extends to south and southwest while redshifted emission extends to north and northeast.  We describe this compact scale outflow separately in Figure~6 as it is slightly shifted with respect to the large scale outflow that we explain below.  Lee \& Ho (2005) also mapped the CO J~=~1-0 emission with a resolution $\sim$10$''$.  In their small scale CO map (Figure~1(b) in Lee \& Ho 2005), there is blueshifted CO emission extending to south-southwest which traces the outflow axis in their interpretation.  And another blueshifted CO emission extending to west-southwest is interpreted as an outflow shell.  Figure 5 in Lee et al. (2000) also showed strong emission extending to south and southwest.  These previous studies revealed that there is a CO outflow component extending to southwest in a close vicinity of RNO91.  All these have same orientation of the blue extended feature seen in our SUBARU and $HST$ images.  Note that the orientation of the redshifted CO emission seen in Arce \& Sargent (2006) also matches well with that of our northeast cavity geometry.

In conjunction with these four features from previous studies, we again suggest that the blue color extended nebulosity seen in our near-infrared images (L, P, and G in Figure 4) might be tracing a root of blueshifted outflow cavity eventually connecting to the CO outflow shell with increasing distance.  The arc-shaped nebulosity (Q1, Q2, and T in Figure~4) extending to the north and to the east through the central source is a counterpart of this cavity.  The redshifted molecular outflow might have constructed the northeast cavity while the blueshifted molecular outflow might have constructed the southwest counterpart.

We then turn our focus on the large scale ($\sim$500$''$) outflow geometry.  Lee et al. (2000) presented the CO J = 1-0 emission associated with RNO91 and found that the molecular outflow has a large scale southeastern lobe.  Their outflow has an wide-opening angle ($\sim$160$^{\circ}$) and the centroid of the outflow emission seems to curve eastward with increasing distance from the source.  They suggested that there was an interaction with ambient material that guides the orientation of the outflow.  Alternatively, it could be that the intrinsic outflow axis is changing with time as a result of precession in the exciting source (Lee et al. 2000).  Precession of the outflow axis is suggested to be occurred in a few objects such as L1157 and L723 (Gueth et al. 1996, Hirano et al. 1998, Zhang et al. 2000, Bachiller et al. 2001).  From this point of view, there could be that a bottom of the blueshifted outflow has a component which has a direction toward southwest in the vicinity of the source which was seen in our near-infrared images and then blueshifted outflow curves eastward with increasing distance from the source.  This possible interpretation is illustrated in Figure~6 as well.  However, it is still hard to discern a real sense of the outflow orientation using our data alone which has a field of view of 20$''$.

\subsection{South Clump and Dark Lane}

The southern clump (S in Figure~4), located 6$\sim$7$''$ south of the central source, gives an appearance of blue ($H-K$ \textless 0.5, Figure~2) arc-like ridge surrounding a red ($H-K$ \textgreater 0.8, Figure~2) inner component.  The location of this southern clump in our $K$ band image coincides well with that of integrated intensity of $^{13}$CO (1-0) emission mapped by Arce \& Sargent (2006).  Since $^{13}$CO (1-0) emission usually traces the dense gas, this southern clump in our image can be interpreted as the circumstellar gas envelope.  In some cases of class II sources, gaseous component that has been entrained by the outflow may form relatively dense shells and clumps at the edge of the outflow lobe around the source (Arce \& Sargent 2006).  Therefore, this southern clump could be residue after outflow clears away the surrounding gas.

  This south clump is separated from the central source by a gap of 2$''$.  Our $H-K$ color image suggests that this dark lane has no additional reddening here and thus that this gap might not be due to shadowing from a disk around RNO91.  Unfortunately there have been no radio data so far which had enough resolution to resolve this dark lane.  We consider that there is an actual break in the mass distribution here.

\section{Summary}

We have obtained sub-arcsecond near-infrared ($JHK$ band) images of a class II source, RNO91, utilizing the near-infrared camera, CIAO, mounted on the SUBARU 8.2 m telescope.  We presented $JHK$ band and optical images, which are the highest resolution images of RNO91 ever taken.  New insights into the immediate circumstellar environment around RNO91 are obtained.  The main conclusions are as follows:
\begin{enumerate}

\item{
The $K$-band image shows bright circumstellar nebulosity detected within $\sim$2$''$~(300AU) around the central source while it is less conspicuous at shorter wavelengths such as $J$ and optical.  The nebula appears to become more isotropic with increasing wavelength.  P.A. and size of this much red color nebulosity in our $H-K$ color image agree with those of the polarization disk previously suggested by Scarrott, Draper, \& Tadhunter (1993).  These data agreement indicate that this bright circumstellar nebulosity region which follows the reddening law might be attributed to a disk-like structure.}\\

\item{
At $J$ and optical wavelengths, several blue knot-like structures are detected around and beyond the bright circumstellar nebulosity.  These blue knots seen in our images are spread over a region comparable in size to the envelope detected in HCO$^{+}$ J~=~1-0 emission ($\sim$10$''$) surrounding RNO91 (Lee \& Ho 2005).  It is natural to interpret that these knotty reflection nebulae located beyond 2$''$ from central source position may represent disintegrating fragments of an infalling envelope.}\\

\item{
Three-color composite image has an appearance of arc-shaped nebulosity extending to the north and to the east through the central source.  On the other end of this arc-shaped structure, the nebula appears to become more extended (2\rlap.{$''$}3 long) to the southwest (P.A.$\sim$225$^{\circ}$) from the central source position in $J$ band and $HST$ optical images.  We interpret the whole structures as roots of bipolar cavity opening to the northeast and southwest directions.  Redshifted molecular outflow might have constructed the northeast cavity while blueshifted molecular outflow might have created the southwest counter part.}\\

\item{
The southern clump, located 6$\sim$7$''$ south of the central source, gives an appearance of blue arc-like ridge surrounding a red inner component.  The location of this southern clump in our $K$ band image coincides well with that of the integrated intensity of $^{13}$CO (1-0) emission mapped by Arce \& Sargent (2006).  This southern clump in our image can be interpreted as the circumstellar gas envelope.  This south clump is separated from the central source by a gap of 2$''$.  No $H-K$ color change near the dark lane suggests that this gap might not be due to shadowing from a disk around RNO91 but an actual break in the mass distribution.}\\

\end{enumerate}

\bigskip

We thank the telescope staffs and operators at the SUBARU Telescope especially Sumiko Harasawa and Dennis Scarla for their assistance.  We are grateful for having fruitful discussions with Michihiro Takami, Ray Furuya, Takayuki Nishikawa, Takuya Fujiyoshi, Takuya Yamashita, Yasushi Nakajima, and Chin-Fei Lee.  We also thank our referee for the constructive comments that have helped to improve this manuscript.  Based on data collected at Subaru Telescope, which is operated by the National Astronomical Observatory of Japan.  The HST data presented in this paper was obtained from the Multimission Archive at the Space Telescope Science Institute (MAST).  Based on observations made with the NASA/ESA Hubble Space Telescope, obtained at the Space Telescope Science Institute, which is operated by the Association of Universities for Research in Astronomy, Inc., under NASA contract NAS 5-26555.  These observations are associated with program 7387 and 8216 conducted by Dr. Stapelfeldt, K. R.(Principal Investigator).  S. Mayama is financially supported by the Japan Society for the Promotion of Science (JSPS) for Young Scientists.

\newpage


\appendix


\newpage

\begin{figure}
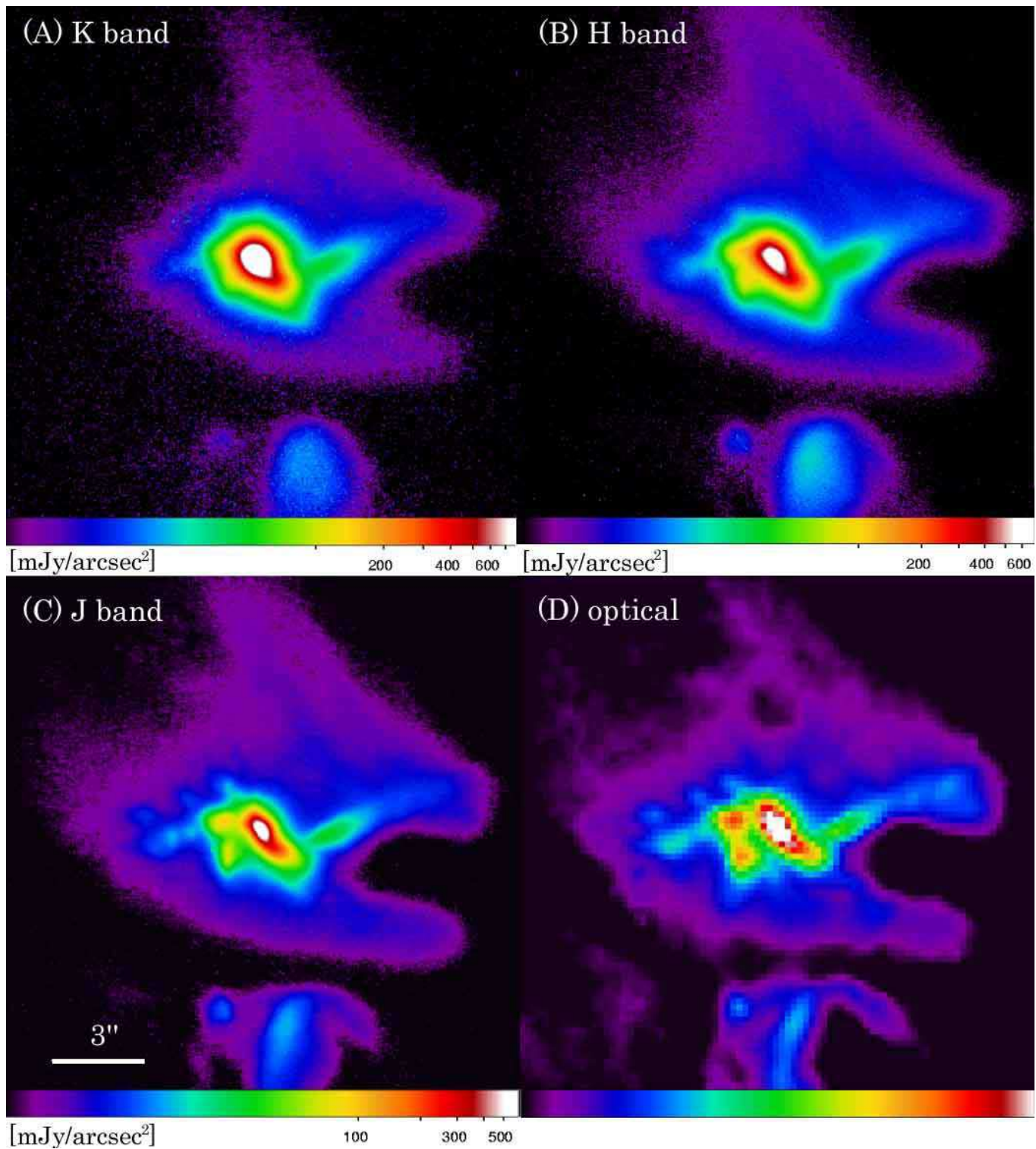

  \begin{center}
    \FigureFile(180mm,180mm){Fig1RonbunWithColorBarForAstroPH.eps}
  \end{center}
  \caption{$J$, $H$ and $K$ band images of RNO91 with an $HST$ optical image (F606W).  Field of View (FOV) is 17$''\times$17$''$.  North is up and east to the left.  }\label{fig:sample}
\end{figure}

\begin{figure}
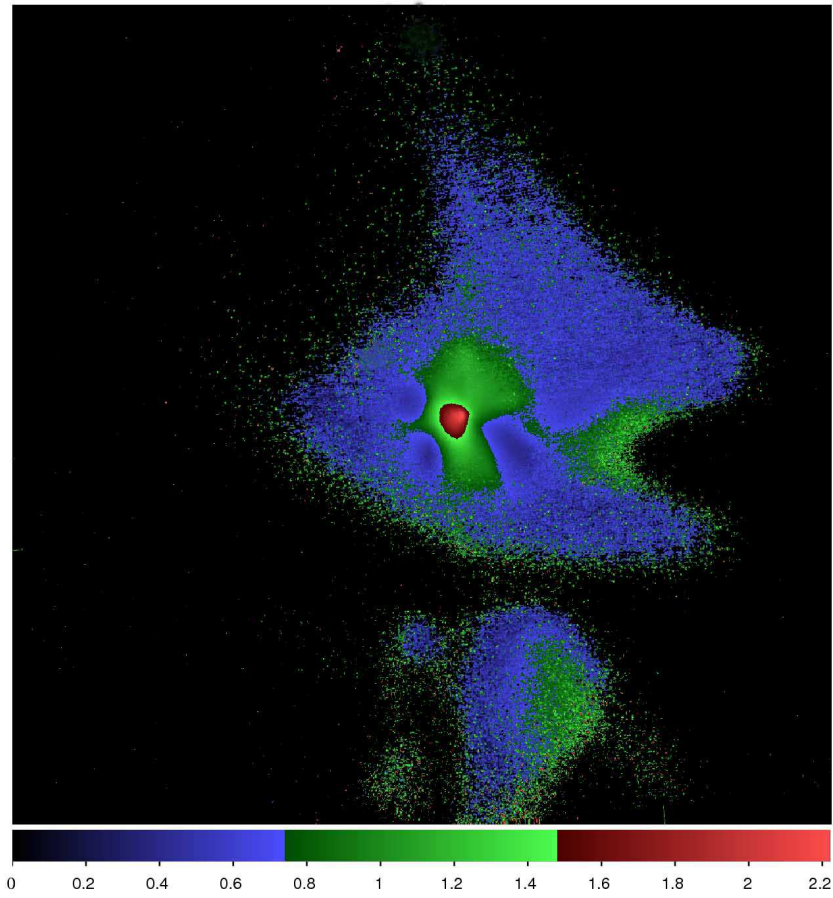

  \begin{center}
    \FigureFile(120mm,120mm){a339Fig2BackBeauty2Color2ForAstroPH.eps}
  \end{center}
  \caption{$H-K$ color image.  FOV is 22$''\times$22$''$.  North is up and east to the left.  Strong emission indicates red color and weak emission indicates blue color.  Unit of color bar is magnitude.}\label{fig:sample}
\end{figure}

\begin{figure}
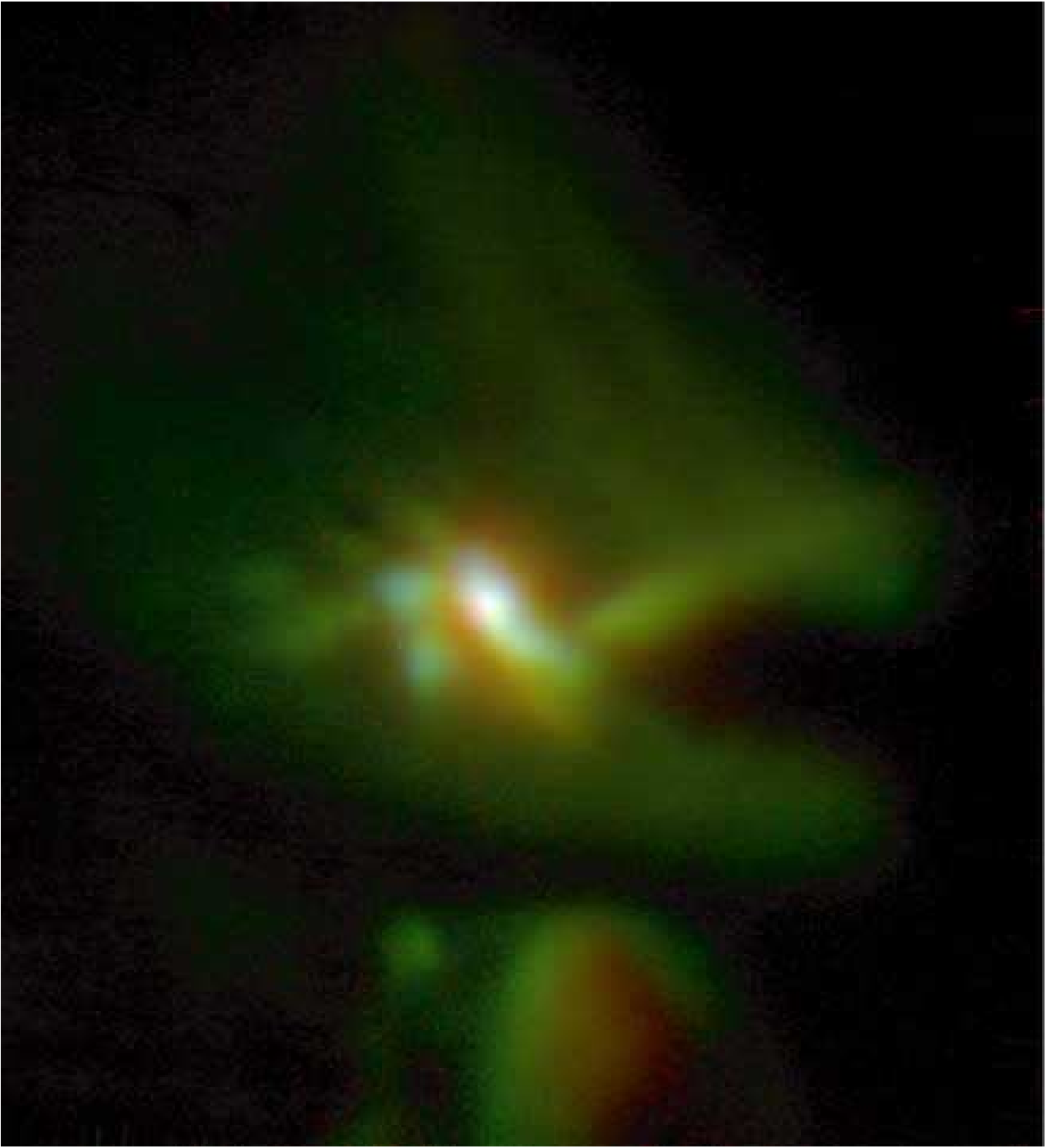

  \begin{center}
   \rotatebox{0}{
    \FigureFile(180mm,180mm){b16Kaitei2-2005CIAOForAstroPH.eps}}
  \end{center}
  \caption{Three-color composite image of RNO91.  The CIAO $K$-band is registered as red, CIAO $J$-band as green, and WFPC2 optical as blue.  North is up and east toward the left.  Note that $HST$ image whose resolution was degraded to the Subaru level was used to make this three-color composite image.}\label{fig:sample}
\end{figure}

\begin{figure}
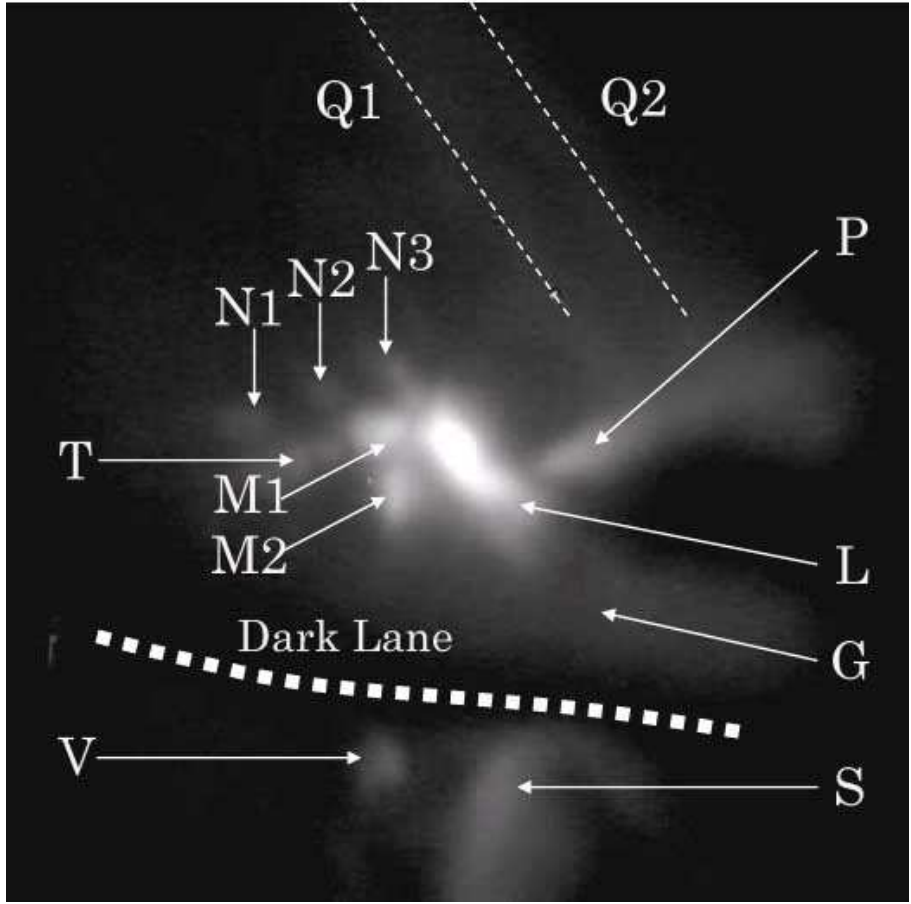

  \begin{center}
    \FigureFile(120mm,120mm){Fig4GreyColor1ForAstroPH.eps}
  \end{center}
  \caption{Identifiers of each structure superimposed on the SUBARU $J$ band image.  Field of view is 17$''\times$17$''$.  M1, M2, N1, N2, and N3 are knot-like structures around the source.  Q1 and Q2 are northern arcs.  T is eastern arc.  P and G are west and southwest arcs, respectively.  S is arc ridge surrounding southern clump.  P, G, and S were initially labeled by Schild, Weir, \& Mathier (1989).}\label{fig:sample}
\end{figure}

\begin{figure}
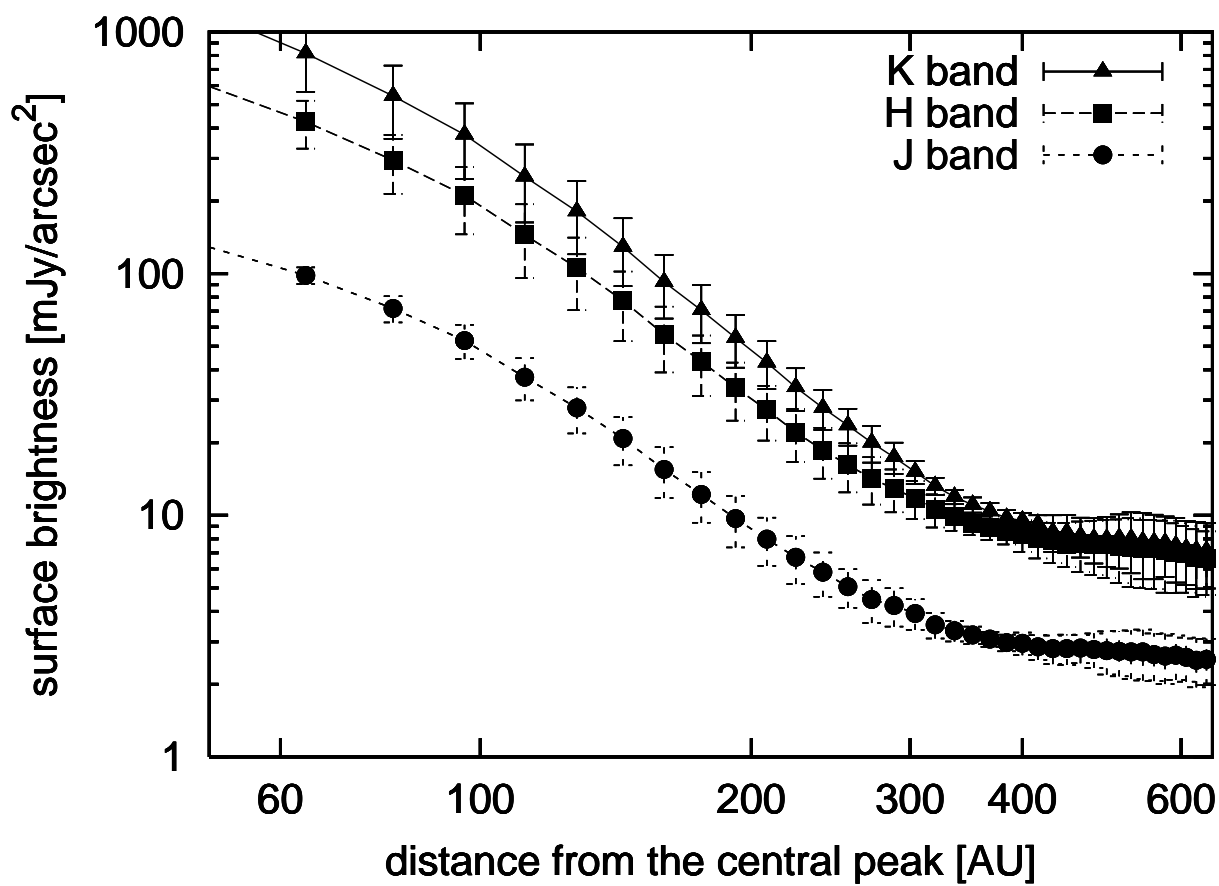

  \begin{center}
   \rotatebox{270}{
    \FigureFile(120mm,170mm){RNO91ProfileFinish1.ps}}
  \end{center}
  \caption{Radial profile of the surface brightness of RNO91 associated nebula in the $JHK$ band.  Statistical error and standard deviation amongst various orientations are calculated to estimate error bars.}\label{fig:sample}
\end{figure}

\begin{figure}
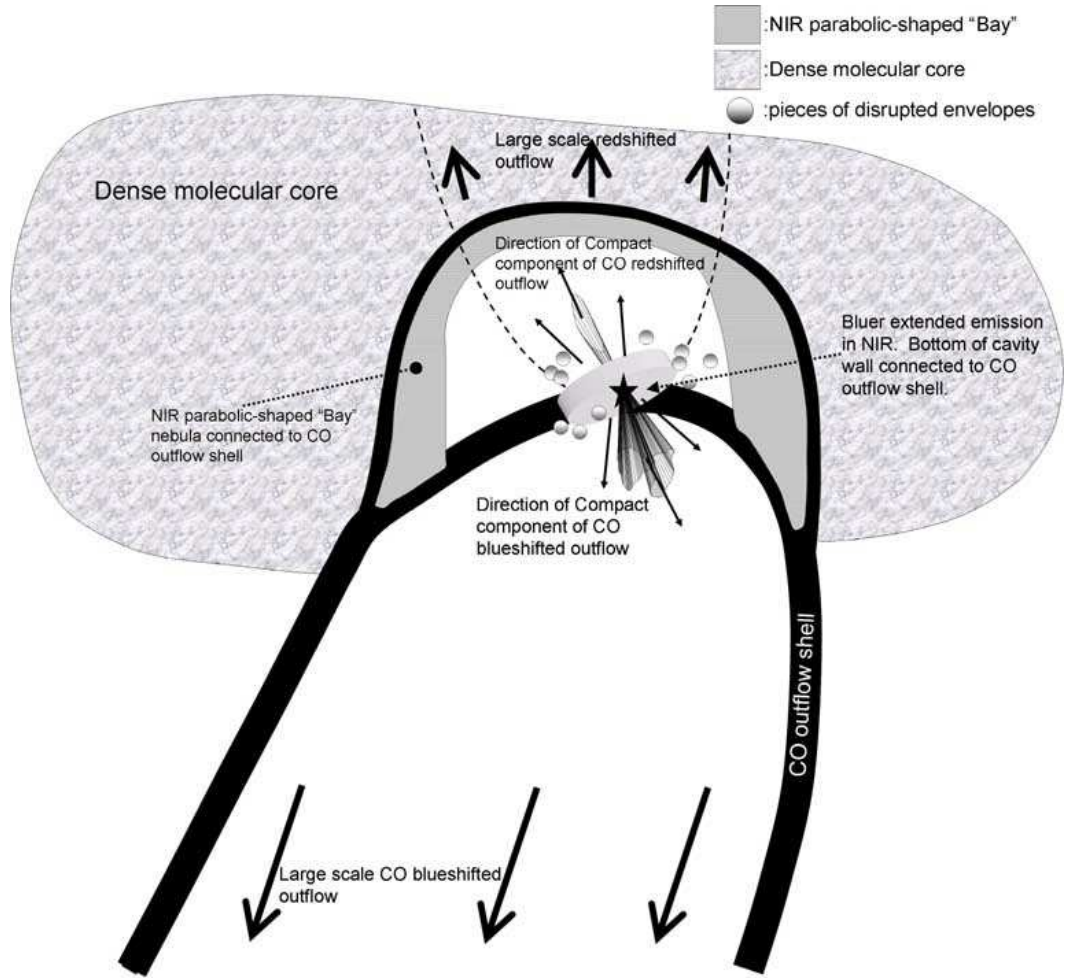

  \begin{center}
    \FigureFile(140mm,140mm){RNO91PontiFig6-3ForAstroPH.eps}
  \end{center}
  \caption{Illustration of front view of RNO91 and its surrounding.  The position of the central source is marked by the star.  The hatched regions represent compact component of molecular outflow resolved by Arce \& Sargent (2006).  The shaded area represents near-infrared parabolic shaped ``Bay$''$ reflection nebulosity seen at optical and near-infrared.  The sizes of disk, outflow, and molecular core are not drawn with the same scale.  The field of view of our SUBARU images covers immediate circumstellar environment around the central source such as disk and roots of bipolar cavity.}\label{fig:sample}
\end{figure}

\end{document}